\documentclass[journal=ancac3,manuscript=article]{achemso}

\makeatletter
\renewcommand\@fnsymbol[1]{\arabic{#1}} 
\makeatother

\usepackage[version=3]{mhchem} 
\usepackage{siunitx}
\DeclareSIUnit{\litre}{l}
\usepackage{xr-hyper}
\usepackage{hyperref}
\usepackage{physics}
\usepackage{braket}
\usepackage{ulem}
\usepackage[font=footnotesize,labelfont=bf]{caption}

\usepackage{etoolbox}

\makeatletter
\renewcommand*{\acs@author@fnsymbol@symbol}[1]{
    \ifcase #1 *\or
    1\or
    2\or
    3\or
    4\or
    5\or
    6\or
    7\or
    8\or
    9\or
    10
    \fi
}
        
\renewcommand*\acs@contact@details{
    {\sffamily *\,E-mail: \acs@email@list }%
    \acs@number@list
}           

\patchcmd{\acs@address@list@auxii}
{\acs@author@fnsymbol{\acs@affil@marker@cnt}}
{\textsuperscript{\acs@author@fnsymbol{\acs@affil@marker@cnt}}}
{}{}

\patchcmd{\acs@address@list@auxii}
{{\acs@author@fnsymbol{\acs@affil@marker@cnt}\@nameuse{@altaffil@\@roman\@tempcnta}\par}}
{{\textsuperscript{\acs@author@fnsymbol{\acs@affil@marker@cnt}}\@nameuse{@altaffil@\@roman\@tempcnta}\par}}
{}{}        

\makeatother 

\author{Andre Pointner}
\affiliation[Institute of Applied Quantum Technologies]{Institute of Applied Quantum Technologies, Friedrich-Alexander-Universität Erlangen-Nürnberg, Erlangen, 91052, Germany}
\email{andre.pointner@fau.de}

\author{Daniela Thalheim}
\affiliation[Experimental Tumor Pathology]{Experimental Tumor Pathology, Universitätsklinikum Erlangen, Friedrich-Alexander-Universität Erlangen-Nürnberg, Erlangen, 91052, Germany}

\author{Sarah Belasi}
\affiliation[Experimental Tumor Pathology]{Experimental Tumor Pathology, Universitätsklinikum Erlangen, Friedrich-Alexander-Universität Erlangen-Nürnberg, Erlangen, 91052, Germany}

\author{Lukas Heinen}
\affiliation[Section of Experimental Oncology and Nanomedicine]{Section of Experimental Oncology and Nanomedicine, Universitätsklinikum Erlangen, Friedrich-Alexander-Universität Erlangen-Nürnberg, Erlangen, 91052, Germany}

\author{Cristian Bonato}
\affiliation[Institute of Photonics and Quantum Sciences]{Institute of Photonics and Quantum Sciences, SUPA, Heriot-Watt University, Edinburgh, EH14 4AS, UK}

\author{Tobias Luehmann}
\affiliation[Applied Quantum Systems]{Applied Quantum Systems, Felix Bloch Institute of Solid State Physics, Leipzig, 04109, Germany}

\author{Jan Meijer}
\affiliation[Applied Quantum Systems]{Applied Quantum Systems, Felix Bloch Institute of Solid State Physics, Leipzig, 04109, Germany}

\author{Rainer Tietze}
\affiliation[Section of Experimental Oncology and Nanomedicine]{Section of Experimental Oncology and Nanomedicine, Universitätsklinikum Erlangen, Friedrich-Alexander-Universität Erlangen-Nürnberg, Erlangen, 91052, Germany}

\author{Christoph Alexiou}
\affiliation[Section of Experimental Oncology and Nanomedicine]{Section of Experimental Oncology and Nanomedicine, Universitätsklinikum Erlangen, Friedrich-Alexander-Universität Erlangen-Nürnberg, Erlangen, 91052, Germany}

\author{Regine Schneider-Stock}
\affiliation[Experimental Tumor Pathology]{Experimental Tumor Pathology, Universitätsklinikum Erlangen, Friedrich-Alexander-Universität Erlangen-Nürnberg, Erlangen, 91052, Germany}

\author{Roland Nagy}
\affiliation[Institute of Applied Quantum Technologies]{Institute of Applied Quantum Technologies, Friedrich-Alexander-Universität Erlangen-Nürnberg, Erlangen, 91052, Germany}
\email{roland.nagy@fau.de}

\title{Optimizing SPION Labeling for Single-Cell Magnetic Microscopy}


\begin{document}

\begin{abstract}

This study explores the correlation between iron mass on cell surfaces and the resultant magnetic field. Human colorectal cancer cells (HT29 line) were labeled with varying concentrations of SPIONs and imaged via a NV center widefield magnetic microscope. To assess the labeling efficacy, a convolutional neural network trained on simulated magnetic dipole data was utilized to reconstruct key labeling parameters on a cell-by-cell basis, including cell diameter, sensor proximity, and the iron mass associated with surface-bound SPIONs.

Our analysis provided quantitative metrics for these parameters across a range of labeling concentrations. The findings indicated that increasing SPION concentration enhances both the cell-surface iron mass and magnetic field strength, demonstrating a saturation effect. This methodology offers a coherent framework for the quantitative, high-throughput characterization of magnetically labeled cells, presenting significant implications for the fields of cell biology and magnetic sensing applications.
\end{abstract}

Traditional methods for studying cell behavior and cell movements, are often limited by their inability to track individual cellular dynamics over extended timescales within life-sustaining environments, due to their reliance on optical accessibility and fluorescent stains \cite{doi:10.1126/science.aaq1067, Chen2018-bp, Roy2002-zk, mi11010034}. Recent advances in imaging technologies that use magnetic fields as information carriers provide a promising solution to overcome these limitations. These approaches exploit the ability of magnetic fields to penetrate biological tissues without interference, benefiting from the inherently low magnetic background of biological systems \cite{C4LC00314D}. Taking advantage of these benefits requires a sensor with high magnetic field sensitivity compatible with biological environments. Current technological developments in imaging and metrology within the field of quantum technologies have brought the NV center in diamond into focus. It provides a promising quantum sensor capable of sensing domains such as nano-nuclear magnetic resonance (Nano-NMR), electric field, temperature, or high precision 2D magnetic field imaging at room temperature \cite{doi:10.1021/acssensors.2c00670, Hollendonner2023-yi, Pogorzelski2024Compact, Kim2019A, Zhang2021Toward, Janitz2022Diamond, Bian2021-pi}. A key issue with respect to magnetic field imaging is the inherent lack of a magnetic signal in biological samples. In recent years, new technologies have emerged to modify cells to magnetically label them. One of the most common methods to magnetically label biological specimens is the use of superparamagnetic iron oxide nanoparticles (SPIONs) \cite{Nasser2024-ie, Friedrich2022-sf}. SPIONs are small, biodegradable, and biocompatible magnetic particles that are employed as magnetic labels that can be attached to specimen or can enter various cell types, including tumor cells \cite{Vangijzegem2023-xn}. The SPIONs are used in medical applications, enabling real-time, non-invasive tracking of labeled cells \cite{Dulinska-Litewka2019-fc}. In cancer research, SPIONs are used to monitor tumor cell migration and interaction with other cells like immune or stem cells within the tumor microenvironment \cite{Mamani2012-xv, Bull2014-ti}. SPIONs can be used for evaluating treatment effects and can be engineered for targeted drug delivery \cite{Zhi2020-lg}. Their magnetic responsiveness and flexibility in functionalization allow them to combine imaging, therapy, and diagnostics in one platform \cite{Wang2017-dm, Fernandez-Barahona2020-zt, Liu2018-as}. 
Labeling with SPIONs has previously been used to detect a variety of biological samples such as individual cells \cite{Glenn2015-gy, Le_Sage2013-at}, tissue slices \cite{Chen2022-cs}, or particle imaging of gastrointestinal tracts of larvae \cite{Mathes2023-ll} with NV centers in diamond. Although this approach has enabled an increased number of applications in recent years, it relies on the challenging detection of faint magnetic fields produced by SPIONs, which can reach magnitudes in the $pT$ to $\mu T$ range within the microenvironment around the SPIONs. In order to use this technology for a wider range of applications, a stronger detectable magnetic field is generally desired to improve both temporal resolution and sensing volume. Given the physical limit to the magnetic moment of SPIONs, an increase in the magnetic field strength of the observed specimen can be achieved by increasing the number of SPIONs attached to the cell surface. While previous studies have focused on applications of this technique, information regarding the iron mass of SPIONs on the cell surface contributing to the magnetic labeling effectiveness is lacking. In this work, we aim to address this open question by investigating the iron mass of SPIONs on the cell surface and its contribution to the magnetic field strength.

\begin{figure}[H]
\centering
\includegraphics{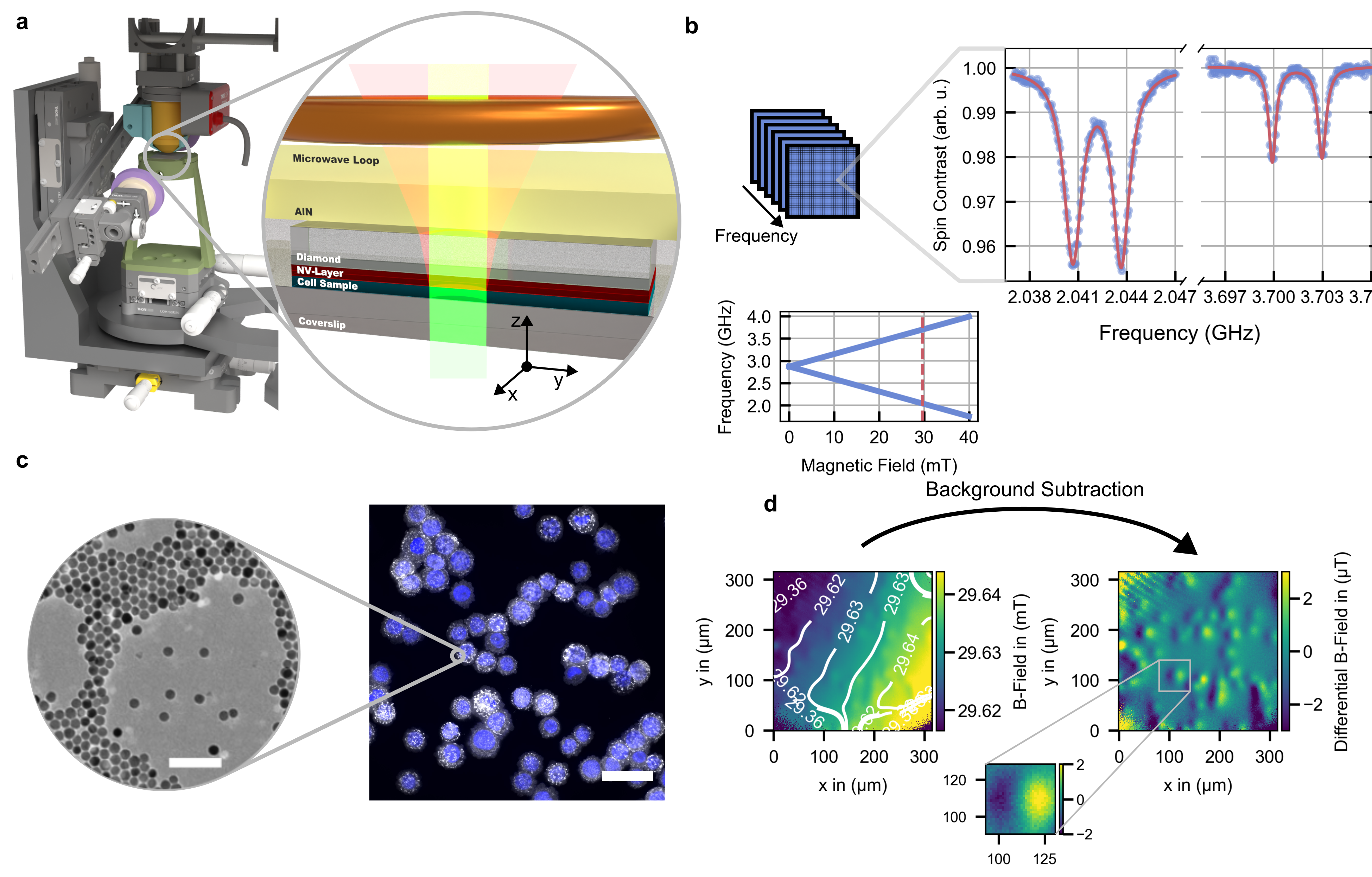}
\caption{NV Center based Widefield Magnetometry. (a) Widefield magnetic field microscope to observe magnetically labeled cells. The sample under investigation is mounted to the NV layer of the diamond substrate with a thin layer of UV curing optical adhesive. The CW-ODMR measurement is performed by exciting the NV centers in the center of the FoV with a \SI{532}{\nm} laser and simultaneously sweeping the frequency of a microwave signal applied to the sample by a copper loop. The resulting PL emission from the \SI{318}{\mu \m} x \SI{318}{\mu\m} large FoV is collected, filtered and projected onto the sensor of an EMCCD camera. (b) Evaluating the electron Zeeman effect observable in the pixel-wise resolved CW-ODMR spectrum reveals the magnetic field projected to the chosen NV axis for each given pixel of the resulting image. (c) Individual HT29 cells magnetically labeled with SPIONs attached to the cell surface. To verify particle binding a fluorescent marker was attached to the SPIONs and fluorescence imaging was performed. The signal displayed in white illustrates the successful binding of the particles as clusters on the cell surface. White scale bar is \SI{50}{\um}. Insert: Transmission Electron Microscopy image of the employed SPION batch to illustrate the size, supplied by the manufacturer\cite{ocean-nanotech}. White scale bar is \SI{100}{\nm}. (d) Removal of the magnetic background created by inhomogeneities in the lab environment, bias magnetic field and microwave antenna reveals the underlying dipole signatures created by the magnetically labeled cells. The expected characteristic shape of the signature, created by the intersection of a magnetic dipole field at an angle respective to the NV axis in the diamond crystal is clearly observable.}\label{fig:1}
\end{figure}

In the presence of an external magnetization field the superparamagnetic core of the SPIONs creates a magnetic dipole field, whose magnetic moment aligns itself with this external field. For the detection of this weak magnetic dipole field, we utilize a NV center based widefield microscopy system, depicted in \autoref{fig:1}a. The system's core component consists of a [100] bulk diamond sample ($>\SI{98}{\percent}$ ${}^{12}$C) with a 10 ppm ${}^{15}N$ nitrogen doping which results in a NV center concentration of \SI{1}{ ppm} within a \SI{1}{\mu\meter}-thick layer (Quantum Diamonds \cite{quantum-diamonds}).  Optical excitation of the NV centers is achieved using a Coherent Verdi \SI{532}{\nm} laser operating at a power of \SI{250}{\milli\watt} measured before the sample. The laser illuminates a field of view (FoV) of 318 x \SI{318}{\um}. The resulting red-shifted fluorescence emission is collected by an air objective (UPLFLN 20x/0.5NA, Olympus) and subsequently undergoes spectral filtering (FELH0600, Thorlabs, Inc.) before detection by a Teledyne Evolve 13 EM-CCD camera. In addition, the widefield microscope contains also a red LED for brightfield images. Spin state manipulation is realized by a loop antenna positioned \SI{2}{\mm} above the diamond surface. A static magnetic field of \SI{30}{\milli\tesla} is set at an angle of approximately \SI{54.74}{\degree} along a NV axis to remove the degeneracy of the $\ket{\pm 1}$ NV center spin states. We chose the crystal axis most practical for our experiment. This magnetic bias field simultaneously magnetizes the SPIONs, which then generate a measurable magnetic field gradient (see \hyperref[suppl:spions]{Supporting Information (SI)}). An overview of the optical and electronic components of the NV center based widefield microscope is detailed in the \hyperref[suppl:exp]{SI}.

To visualize the magnetic fields generated by magnetized SPIONs, we implement continuous-wave optically detected magnetic resonance (CW-ODMR). Each pixel of the camera sensor corresponds to fluorescence emission from a specific region within the FoV in the NV center layer. During continuous laser excitation, an AC-magnetic field delivered via the loop antenna and synchronized with the camera's frame acquisition sequence, enables spatially resolved interrogation of the NV center spin states across the microscope's FoV. The resonance frequencies of the spin states are determined by sweeping the AC magnetic field frequency across the corresponding $\ket{0} \rightarrow \ket{\pm 1}$ transitions. The resonance frequency is identified by a spin state transfer from the ($\ket{0}$) to the ($\ket{\pm 1}$) states, producing a characteristic contrast within the ODMR spectrum of each pixel, as illustrated in \autoref{fig:1}b. The magnetic field strength along the interrogated NV axis for every pixel is determined by evaluating the electron Zeeman splitting $\Delta f=2\gamma B_z$ observable by extracting the center of the hyperfine doublets. In order to generate a magnetic field map we capture 300 sets of two images each per frequency step: one with the AC-magnetic field active and one without, resulting in a spin contrast image. Our magnetic field microscope achieves a magnetic field sensitivity of up to \SI{200}{\nano\tesla\Hz\tothe{-\frac12}} per pixel or a volume-normalized sensitivity of \SI{254}{\nano\tesla \mu\meter\tothe{\frac32} \Hz\tothe{-\frac12}}, ultimately limited by the available optical excitation power at the NV layer (see \hyperref[suppl:sens]{SI}).

In order to characterize the iron mass on cells and thus assess the labeling effectiveness, we used human colorectal cancer cells (HT29 line). They were prepared for magnetic measurements via a surface-targeted labeling strategy using sequential \SI{30}{\min} incubations on living cells in RPMI-1640 medium: first with a primary anti-EpCAM antibody, then a biotinylated secondary antibody, and finally with \SI{20}{\nm} core diameter, streptavidin-coated SPIONs (SHS-20, Ocean Nanotech LLC \cite{ocean-nanotech}, more  details are given in the SI).

We prepared various concentrations of the SPION solution (\SI{1}{\milli\gram\per\milli\liter}) by diluting it with DI water (1:50 SPION Solution/DI, 1:25 SPION Solution/DI, 1:10 SPION Solution/DI, 1:5 SPION Solution/DI and 1:1 SPION Solution/DI) for the magnetic field measurements, in order to produce varying labeling strengths. The labeled cells were subsequently immobilized on coverslips by cytocentrifugation and fixed with paraformaldehyde. This produces a somewhat homogeneously spread distribution of individual cells across the microscope coverslip. Successful cell surface labeling was confirmed by fluorescence microscopy, visualizing the SPIONs indirectly via Cy5-biotin binding and counterstaining nuclei with DAPI staining, prior to mounting for magnetic measurements as depicted in \autoref{fig:1}c (see \hyperref[suppl:prep]{SI} for more a more detailed description of the sample preparation process).

After sample preparation, the cells are mounted on the NV center layer of the diamond. Due to the inverted geometry of our experiment the sample has to be fixed reversibly to the diamond. To provide a higher throughput, we opted to glue the sample with an optical adhesive to the diamond surface instead of centrifuging the cells directly on the diamond surface. This is in part done because the centrifuged cells tend to peel off of the surface and the optical adhesive can be fully removed via chemical cleaning. To mitigate sample tilt and stand-off introduced by the viscosity of the adhesive, we used a home-built mechanism to gently press the diamond and the cell sample together with a controlled pressure to mitigate variations in the resulting cell distance to the sensing layer between sample runs. After pixel-wise evaluation of the magnetic field strength across the FoV, we subtract the quadratic magnetic background resulting from inhomogeneities in the bias field, as illustrated in \autoref{fig:1}d. Removal of the background field $B_0$ reveals the distinct magnetic dipole signature of individual cells labeled with SPIONs, as seen in the inset of \autoref{fig:1}d.

\begin{figure}[H]
\centering
\includegraphics{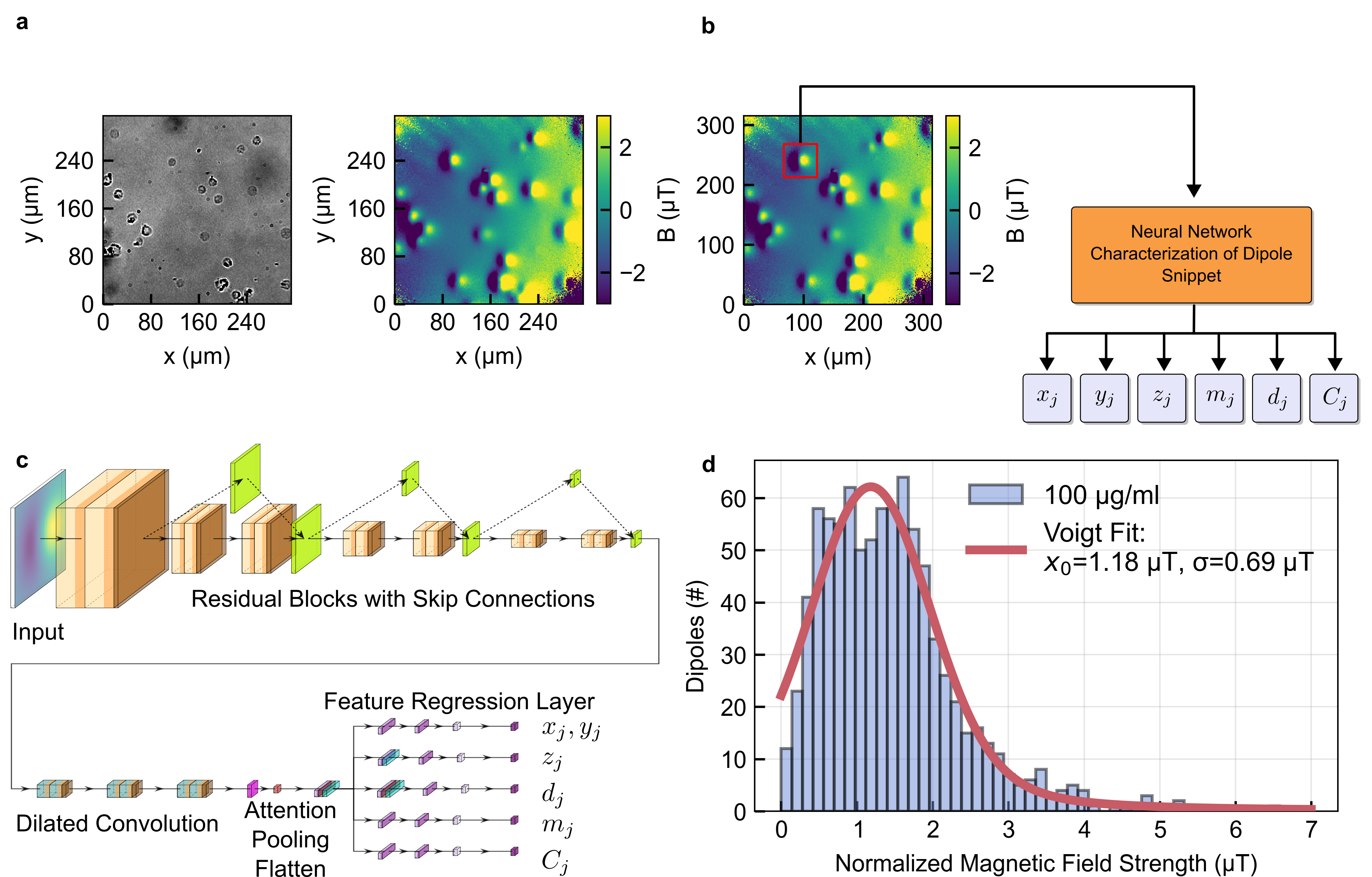}
\caption{Field strength normalization of magnetically labeled cells. (a) Brightfield microscopy image showing circular cell structures. Small black spots are dust accumulation on the camera sensor. Corresponding magnetic field map displaying dipole signatures with color-coded magnitude or the reconstructed magnetic field strength along the chosen NV axis. (b) Schematic representation of the dipole snippet selection process and neural network classification pathway. Selected regions of interest are processed through a convolutional neural network to characterize magnetic dipole properties such as position $(x_j, y_j)$ within the snippet, distance $z_j$ to the NV layer, iron mass $m_j$ distributed on the cell surface and cell size $d_j$. A constant background $C_j$ is added as a parameter to allow the network to compensate for imperfect background subtractions. (c) Architecture of the employed Neural network. Convolutional layers with residual skip connections encode the spatial information of the dipole snippet. Dilated convolutions, attention, and pooling layers condense the encoded information. A subsequent flattening into a regression layer with a multi-head feature interpretation transforms the input snippets (32×32×1) into quantitative dipole parameters. (d) Histogram of magnetic field strength normalized by simulation of the networks prediction for the measured dipoles. A Voigt profile is fit to the data with a peak position of \SI{1.18}{\mu\tesla} and a width of \SI{0.69}{\mu\tesla} for samples prepared with \SI{100}{\mu\g\per\milli\litre} SPION concentration. This normalization approach enables standardized quantification of cellular magnetic labeling efficiency across different experimental conditions.}
\label{fig:2}
\end{figure}

The magnetic field map, which reveals the magnetic dipole field characteristics of individually labeled cells, also indicates their spatial positions. Each signature follows a distinct pattern characterized by the cross-section of the magnetic dipole field in the NV center layer of the diamond. For each cell a magnetic field strength can be assigned, defined as the peak value of the magnetic dipole field from all SPIONs on the cell surface after subtraction of the bias field. This bijective relationship between a single cell and its magnetic signature is shown in \autoref{fig:2}a. To evaluate the iron mass on cells, it is necessary to measure the magnetic field strength of each cell, resulting from the cell diameter, iron mass, and the cell's distance to the NV center layer. The challenge lies in the varying distance between the cells and the NV center layer due to the viscosity of the adhesive, both within a single sample as well as between multiple samples. A \SI{1}{\degree} tilt of the cell cover slide relative to the NV Centers would introduce an offset up to \SI{17.5}{\mu\m} across the \SI{1}{\milli\m} sample area and therefore in different magnetic field measurement results. Additionally, the distinct shapes of the magnetic dipoles in \autoref{fig:2}a depend on the amount of SPIONs on the cell surface, the distance between the cell and the NV center layer, but also the cell size. These dependencies imply that evaluating the magnetic properties of cells based on the iron mass is complex. To address this issue, a convolution neural network is utilized to normalize the measured magnetic field data and determine the iron mass for each cell.

Quantifying the iron mass on individual cells from their magnetic field presents a significant inverse problem. The measured field, $\mathbf{B}(\mathbf{r})$, is a superposition of dipole fields from all SPIONs attached to the cell surface: \begin{align} \mathbf{B}(\mathbf{r}) = \sum_k \mathbf{B}_k(\mathbf{r})\text{,} \label{eq:2} \end{align} making direct reconstruction of individual dipole parameters or normalization of the measurement results computationally intractable. The forward problem is a function of distance, iron mass on the cell surface, and the size of the cell, as the dipole orientation is given by the externally applied magnetic field. While inverting the problem is not trivial, the clear physical relationship is distinctive. Since these parameters are encoded in the magnitude and spatial extent of the measured magnetic dipole, we can leverage the spatial resolution ($\approx\SI{1.3}{\mu\meter}$ pixel size) of the NV center microscope in addition to the scalar magnetic field measurements taken for each pixel of the camera sensor. Though magnetic field vector reconstruction is possible with NV centers, it would either require a larger experimental overhead in realigning the magnetic bias field to different NV axis or would reduce the scalar field sensitivity, by distributing the fluorescence contrast on multiple NV center orientations \cite{Steinert2010-nv}. For this reason, we chose a scalar magnetic field reconstruction to characterize the dipole signal. 

We employ a convolutional neural network (CNN) architecture to analyze the magnetic dipole field signature from a single cell, predicting the underlying physical quantities: the distance to the NV center layer, cell size, and iron mass on the cell surface. As illustrated in (\autoref{fig:2}b), we segment the magnetic field map into smaller snippets corresponding to the lateral position of each footprint. By segmenting it into a constant $32\times 32$ pixel size (chosen by experimental observation to include most of the dipole information), we provide a normalized input magnetic dipole field segment to the neural network. This input snippet is fed into a custom-built deep learning network (illustrated in \autoref{fig:2}c) comprising several convolutional layers to encode features in the input image. Residual skip connections allow information to bypass some of the deeper convolutional layers, a self-attention layer enables the network to focus on distinct features of the magnetic dipole footprint. A final regression network interprets the output of the convolutional block with respective feature heads and converges towards the six parameters required for a full reconstruction of the SPION-labeled cell within the image segment, namely the position ($x_j, y_j$) within the segment, the distance $z_j$ to the NV center layer, cell diameter $d_j$, iron mass $m_j$ on the cell surface, and a homogeneous background $C_j$ to allow a more robust reconstruction of the experimental ground truth (see \hyperref[suppl:NN]{SI} for details). The output parameters allow a detailed analysis on the SPION labeling efficiency. 

Precise \textit{a priori} knowledge of the cell distance on a cell-by-cell basis is not available for the investigated samples, as the viscosity of the optical adhesive and tilt of the sample slide to the diamond surface cause variations in the cell-to-diamond distance. Additionally, precise information on the iron mass is unknown and could be estimated on a sample wide basis at best (see \hyperref[suppl:iron]{SI}). We therefore turned to numerical simulations for the training data, in order to generate large amounts of magnetic field snippets with a known ground truth. Given that SPIONs tend to cluster on the cell surface, as observed in fluorescence images of the labeled cells (\autoref{fig:1}c), we distribute $N$ points on a spherical surface with diameter $d_j$, representing a single cell. Each point is modeled as a perfect magnetic dipole whose magnetic moment is comprised of a number of SPIONs that is computed by distributing a iron mass $m_j$ across the $N$ points on the spherical surface. This averaging is valid, as the resulting magnetic moment of a magnetically saturated cluster of identical SPIONs scales linearly with the number of particles in the volume:
\begin{align}
    \mathbf{m}_C = \mathbf{M}_CV_C = N\mathbf{m}_PL\bigg(\frac{\mu_0\mathbf{H}\mathbf{m}_P}{k_BT}\bigg) \approx N\mathbf{m}_P\text{,}
\end{align}
where $\mathbf{M}_C$ is the total magnetization of the cluster, $V_C$ is the volume of the cluster, $N$ is the number of particles in the volume and $\mathbf{m}_P$ the magnetic moment of the individual particle, $\mathbf{H}$ the magnetization field, $\mu_0$ the vacuum permeability, $k_B$ the boltzmann constant and $T$ the temperature, the Langevin function $L$ approaches 1 for saturation magnetization. A detailed derivation is illustrated in the \hyperref[suppl:sim]{SI}. The employed SPIONs (\SI{20}{\nm} core from Ocean NanoTech, \#SHS-20) observe a reported magnetic moment of \SI{8.6e-19}{\ampere\metre\squared} for a bias field of \SI{40}{\milli\tesla} \cite{Glenn2015-gy}. Since our bias field strength of \SI{30}{\milli\tesla} approaches saturation \cite{ocean-nanotech} ($\Rightarrow L\rightarrow 1$), we can confidently simulate the experimental conditions with the presented approach. The validity of this approach and the magnetic properties inferred from the work by Glenn \textit{et al.} (2015) is qualitatively confirmed, by comparing the simulation results with measurements of the iron mass per cell of control groups of the measured samples (see \hyperref[suppl:fig:iron]{SI}).

Adding experimental details, such as dipole overlap and noise similar to the observed values to the simulation data, provides a training set robust enough (see \hyperref[suppl:sim]{SI} for details on the simulation and the training process) to train a neural network for application on experimental data.

As the neural network provides estimations for the cell distance $z_j$, the iron mass $m_j$ and the cell size $d_j$, we can use the network prediction to create a simulative reconstruction of the analyzed magnetic field snippet. This fact can be used to quantify the SPION labeling process, since we are able to normalize the magnetic field strength created by this technique, regardless of distance to the sensing layer or cell size, since we can fix the distance and cell size for this reconstruction and use the predicted iron mass as the reference metric. As shown in \autoref{fig:2}d, the expected normalized magnetic field strength of a single SPION labeled cell for a cell-diamond distance of \SI{15}{\mu\meter} and a cell size of \SI{12}{\mu\meter} is approximately normal distributed and observes a distinct mean around a magnetic field amplitude of \SI{1.34}{\mu\tesla} for a SPION concentration of \SI{100}{\mu\g\per\milli\litre} during labeling.

\begin{figure}
\centering
\includegraphics{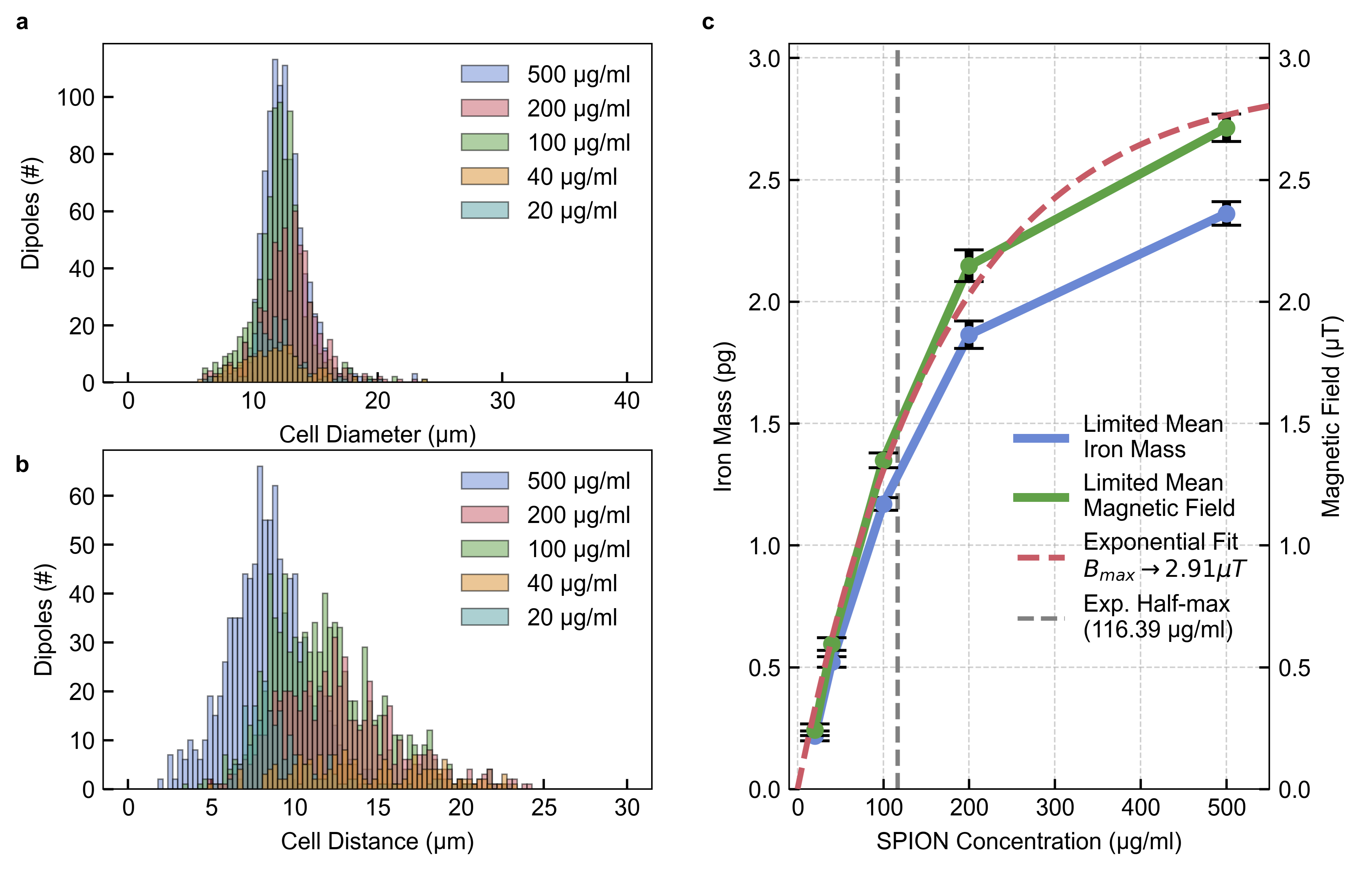}
\caption{
Evaluation of the magnetic labeling efficiency with respect to the SPION concentration based on NN predictions. (a) Cell diameter distribution prediction for all observed samples. A distribution centered around a mean of approximately \SI{12}{\mu \m} is observed. This matches visual confirmation in brightfield images taken prior to a magnetic field measurement (compare \autoref{fig:2}a). (b) Cell distance distribution for all observed samples. A clear shift between an older highly viscous adhesive resulting in a larger distance between samples and NV layer and a fresh batch of adhesive with a lower viscosity used in the \SI{500}{\mu\gram\per\milli\liter} illustrate the problem of magnetic field label normalization, as the measured magnetic field strenths are expected to be much higher, compared to the more distant samples. This variation is visible even between samples of the same batch, due to variance in the mounting process. (c) Reconstructed iron mass distribution and magnetic field trend of the normalized cell signatures. The blue graph displays the predicted mean iron mass of all cells over the applied SPION concentrations. An expected increase in iron mass on the cells is observed and is in line with the naturally resulting increase in maximum magnetic field strength per cell. The green graph illustrates the simulative normalization of magnetic field strength for a cell with \SI{12}{\mu\meter} diameter at a distance of \SI{15}{\mu\meter} with the corresponding iron mass distributed across the cell surface.  Simulating the magnetic field with our simulation framework for a fixed distance and cell diameter allows semi-normalization of the magnetic field strengths expected by a given SPION concentration. The maximum magnetic field strength average we achieved for this normalization cell reaches \SI{2.71}{\mu\tesla} and a potential maximum achievable magnetic field saturates to \SI{2.91}{\mu\tesla} for higher SPION concentrations. A cost-benefit trade-off is concluded at the half maximum which is reached at a SPION concentration of \SI{116.39}{\mu\gram\per\milli\liter}, as illustrated by the dashed grey line.}
\label{fig:3}
\end{figure}

A comprehensive characterization of labeled cells enables a quantitative comparison of labeling effectiveness as a function of SPION concentration (see \hyperref[suppl:prep]{SI} for experimental details). The network's predictions for cell diameter and separation distance from the NV layer fall within a range amenable to experimental verification. As can be anticipated, the cell diameter, derived from a homogeneous cell line, exhibits reasonable variation across all samples, as illustrated in \autoref{fig:3}a. The variations can be explained by differences in general cell vitality (sickly or destroyed cells appear much larger), cytospin centrifugation of the cells on the cover slide, and an applied pressure from the mounting process. In contrast, the cell-NV center separation distance (displayed in \autoref{fig:3}b) is inherently more variable, reflecting the combined influence of cell morphology, adhesive viscosity, and subtle variations in the mounting procedure such as tilt relative to the diamond. Statistically, we expect the average cell-NV center separation to approximate the cell thickness plus an additional offset attributable to the adhesive's surface tension. While our measurements generally reflect this expectation, a noticeable bias variation is observed in smaller sample sets, highlighting the inherent challenges in achieving consistent mounting of the samples to the diamond. Especially the most recent measurements with the highest SPION concentration of \SI{500}{\mu\gram\per\milli\liter} show a significant shift from the other concentrations. This is attributed to the batch of optical adhesive used, that was purchased in the later stages of the experimental runs, providing an adhesive with a lower viscosity, as the previously used adhesive, resulting in a closer mean distance of the cells from the NV layer.  While these inconsistencies would make a labeling evaluation between multiple samples impossible, our machine learning approach compensates these variations.

Beyond cellular morphology, our methodology provides an estimate of the iron mass contributing to the magnetic dipole field associated with the cell-surface labeling. This capability enables a direct, quantitative comparison of the labeling process under varying conditions, thereby assessing its effectiveness. A priori, one might expect an increased SPION concentration would yield to an increase in the detected magnetic field strength, up to a saturation point dictated by the density of available surface receptors. Our measurements (\autoref{fig:3}c) show, that an application of as much as \SI{500}{\mu\g\per\milli\liter} still shows an increase in the magnetic field observed, however saturation is clearly observable in our measurements. This concludes that the EpCAM receptors are not yet fully saturated. The predicted mean achievable magnetic field strength for the normalized \SI{12}{\mu\meter} diameter cells at \SI{15}{\mu\meter} distance from the sensor for the highest SPION concentration reaches \SI{2.71}{\mu\tesla}. This labeling strength allows reliable detection (SNR > 6) of an individual cell at a distance of up to \SI{20}{\mu\meter} (see \hyperref[suppl:sens]{SI} for more details). Extrapolating the data shows a achievable maximum that approaches \SI{2.91}{\mu\tesla}. The half-maximum value displays a good trade-off between the amount of SPIONS required and the achieved magnetic field and is reached at a SPION concentration of \SI{116.39}{\mu\gram\per\milli\liter}. Though deviations from this value are to be expected due to possible inhomogeneous distribution of particles or a varying EpCAM density on the cell surface. The predicted mean iron masses align well with experimentally determined iron contents within prepared samples (see \hyperref[suppl:iron]{SI}). The presented results allow us to make a confident assessment of the magnetic labeling process and allow us to determine an optimum for the applied iron mass on the cell surfaces with respect to the achievable magnetic labeling strength.

In summary, we have demonstrated a novel approach for the quantitative characterization of magnetically labeled cells using a combination of wide-field magnetic microscopy and deep learning. Our network-based analysis assesses cellular parameters, including diameter and, crucially, the separation distance from the sensor, a parameter often challenging to ascertain with conventional techniques. Furthermore, the method provides a direct estimate of the iron mass associated with cell-surface labeling, enabling quantitative assessment of labeling efficiency dependent on SPION concentration. The observed saturation in SPION load establishes a physical limit on the magnetic labeling strength achievable per cell, constrained by the density of available linkers on the cell surface. Our results quantify the cellular magnetic signature as a function of SPION load approaching this limit. The presented combination of HT29 cells and the employed SPIONs allows a reliable detection of indivual cells up to \SI{20}{\mu\meter} away from the sensor. Exceeding this distance requires exceeding the saturation threshold, which necessitates either employing SPIONs with larger intrinsic magnetic moments, or enhanced labeling strength through the usage of additional cell adhesion target systems, such as EGFR or HER2. This would allow exceeding the EpCAM limited receptor density on the cell surface.

\paragraph{Supporting Information}
Discussion of Magnetic Properties of Superparamagnetic Iron Oxide Nanoparticles, Experimental Details, Magnetic Field Sensitivity of CW-ODMR Experiments, Unnormalized Magnetic Field Data, Magnetic Field Distribution Discussion, Simulation Details, Neural Network Training Description, Sample Preparation Details, Reference MP-AES Measurements

\begin{acknowledgement}
Roland Nagy was supported by the Deutsche Forschungsgemeinschaft (DFG) NA1764/2-1 and INST 90/1252-1 FUGG, as well as BMBF (QUBIS). Regine Schneider-Stock was supported by the Deutsche Forschungsgemeinschaft (DFG) SCHN477/19-1 and BMBF (QUBIS). Christian Bonato was supported by the Engineering and Physical Sciences Research Council (EPSRC), through the UK Quantum Hub for Biomedical Research (Q-BIOMED, EP/Z533191/1). 
\end{acknowledgement}

\bibliography{spion_optimization_manuscript}

\end{document}